\documentclass[preprint,12pt,authoryear]{elsarticle}

\usepackage{amssymb}
\usepackage{lineno}
\usepackage{booktabs}
\usepackage{bm}
\usepackage{multirow}
\usepackage[table,xcdraw]{xcolor}
\usepackage{xr-hyper}
\usepackage{hyperref}

\newcommand{\mone}{CCLM4\mbox{-}CanESM2}
\newcommand{\mtwo}{CCLM4\mbox{-}MIROC5}
\newcommand{\mthree}{RCA4\mbox{-}MPI\mbox{-}ESM\mbox{-}LR}
\newcommand{\mfour}{RCA4\mbox{-}NorESM1\mbox{-}M}
\newcommand{\mfive}{RCA4\mbox{-}CNRM\mbox{-}CM5}
\newcommand{\msix}{RCA4\mbox{-}IPSL\mbox{-}CM5A\mbox{-}MR}
\newcommand{\mseven}{RCA4\mbox{-}HadGEM2\mbox{-}ES}
\newcommand{\meight}{RCA4\mbox{-}EC\mbox{-}EARTH}
\newcommand{\mnine}{HIRHAM5\mbox{-}EC\mbox{-}EARTH} 
\newcommand{\mten}{HIRHAM5\mbox{-}NorESM1\mbox{-}M}
\newcommand{\melev}{HIRHAM5\mbox{-}HadGEM2\mbox{-}ES}
\newcommand{\mtwel}{HIRHAM5\mbox{-}MPI\mbox{-}ESM\mbox{-}LR}
\newcommand{\mthirt}{HIRHAM5\mbox{-}CNRM\mbox{-}CM5}
\newcommand{\mfourt}{HIRHAM5\mbox{-}IPSL\mbox{-}CM5A\mbox{-}MR}
\newcommand{\mfivet}{COSMO\mbox{-}crCLIM1\mbox{-}EC\mbox{-}EARTH}
\newcommand{\msixt}{COSMO\mbox{-}crCLIM1\mbox{-}NorESM1\mbox{-}M}
\newcommand{\msevent}{COSMO\mbox{-}crCLIM1\mbox{-}HadGEM2\mbox{-}ES}
\newcommand{\meighteen}{RACMO22E\mbox{-}CNRM\mbox{-}CM5}
\newcommand{\mnineteen}{RACMO22E\mbox{-}EC\mbox{-}EARTH}
\newcommand{\mtwenty}{RACMO22E\mbox{-}IPSL\mbox{-}CM5A\mbox{-}MR}
\newcommand{\mtwentyone}{RACMO22E\mbox{-}HadGEM2\mbox{-}ES}

\journal{}

\begin{document}
\begin{frontmatter}

\title{Assessing Future Wind Energy Potential under Climate Change: The Critical Role of Multi-Model Ensembles in Robustness Assessment}

\author[DICCA]{Andrea Lira-Loarca} \ead{andrea.lira.loarca@unige.it} 
\author[DICCA,INFN]{Francesco Ferrari\corref{cor}} \ead{francesco.ferrari@.unige.it}
\author[DICCA,INFN]{Andrea Mazzino} \ead{andrea.mazzino@unige.it}

\affiliation[DICCA]{organization={Department of Civil, Chemical and Environmental Engineering. University of Genoa},
            addressline={Via Montallegro 1}, 
            city={Genoa},
            postcode={16145}, 
            state={Genoa},
            country={Italy}}
\affiliation[INFN]{
            organization={Istituto Nazionale di Fisica Nucleare, Sezione di Genova},
            addressline={Via Dodecaneso 33}, 
            city={Genoa},
            postcode={16146}, 
            state={Genoa},
            country={Italy}}
\cortext[cor]{Corresponding author}

\begin{abstract}
Accurate projections of wind energy potential under climate change are critical for effective long-term energy planning. While previous studies have highlighted the value of multi-model ensembles, they often fall short in capturing the full spectrum of uncertainties and temporal dynamics relevant to wind resource reliability. This paper presents one of the most comprehensive assessments to date, leveraging a large ensemble of 21 high-resolution RCM-GCM combinations from the EURO-CORDEX initiative to evaluate future wind energy conditions across Europe under the RCP8.5 scenario. Moving beyond mean values, we incorporate a novel event-based framework to analyze persistent high- and low-wind episodes using ERA5-derived percentile thresholds --capturing operationally critical conditions that influence turbine performance and grid stability. To ensure statistical rigor, we apply the IPCC AR6 `Approach C' for robustness assessment, distinguishing climate signals from internal variability and quantifying model agreement. Crucially, we demonstrate that projections based on limited sub-ensembles can lead to contradictory or misleading conclusions, underscoring the essential role of ensemble diversity. The combination of spatial granularity, temporal detail, and formal uncertainty quantification makes this study a significant advancement in climate-informed wind energy research and a valuable tool for resilient energy system design.
\end{abstract}

\begin{highlights}
\item A 21-member EURO-CORDEX ensemble is used to assess future wind energy trends in Europe under RCP8.5.
\item No clear or consistent climate change signal is found across the full ensemble for future wind power potential.
\item Smaller model subsets yield conflicting results, highlighting risks of under-sampling.
\item A formal robustness assessment (IPCC AR6 Approach C) reveals key spatial patterns of change and uncertainty.
\end{highlights}


\end{frontmatter}


\section{Introduction}
The global transition toward renewable energy sources has become a critical priority in the face of intensifying climate change, environmental degradation, and the imperative for sustainable development \citep{Bogdanov2021}. Wind energy, in particular, has emerged as a cornerstone of this transition due to its technological maturity, low environmental impact, and increasingly competitive costs \citep{Jansen2020}. The Intergovernmental Panel on Climate Change has emphasized that wind energy, along with other low-emission technologies, plays a crucial role in achieving deep decarbonization across sectors and is increasingly viable, both technically and economically, as a mitigation strategy \citep{IPCC2023}. Its widespread adoption is also driven by the scalability of wind technologies, their applicability across diverse geographies, and their capacity to integrate into existing power grids with increasing efficiency \citep{Yao2020}.

However, planning future wind energy infrastructure requires a robust understanding of how wind resources may evolve under different climate scenarios. Wind is inherently variable and influenced by complex atmospheric dynamics, including large-scale circulation patterns, local topography, boundary-layer processes, and land-sea interactions. Climate change introduces additional layers of complexity, with the potential to significantly alter wind regimes across spatial and temporal scales. Shifts in temperature gradients, storm tracks, and jet streams may all impact the distribution and intensity of wind speeds, thereby affecting the viability and efficiency of wind energy systems \citep{Pryor2020, Karnauskas2018}. These changes are especially critical given the long investment horizons and infrastructure lifespans associated with wind energy development, making accurate projections of future wind conditions indispensable for informed decision-making \citep{Pryor2020, Solaun2020}.

Climate models have become essential tools for understanding future atmospheric behavior and for projecting wind resource availability under various greenhouse gas emissions pathways \citep{jung2022windreview}. These models simulate interactions between the atmosphere, ocean, cryosphere, and land surface, offering valuable insights into the potential trajectories of future climates. Nevertheless, every climate model is characterized by structural uncertainties arising from differences in spatial resolution, parameterizations of physical processes, and initial condition assumptions. These discrepancies can lead to divergent outputs even under identical forcing scenarios, limiting the reliability of projections derived from individual models \citep{Knutti2010,Cael2022, Tett2022}.

To overcome these limitations, the climate science community has increasingly adopted multi-model ensemble approaches. By integrating results from multiple models, ensemble projections provide a more comprehensive depiction of the range of plausible future outcomes and allow for a probabilistic treatment of uncertainty \citep{Tebaldi2007,Lewis2017}. The use of ensembles is particularly well-suited to wind energy applications, where both the mean state and the variability of wind fields are of critical importance. Ensembles can account for uncertainties in emissions scenarios, model physics, and internal climate variability, leading to more robust and policy-relevant assessments \citep{Deser2012, Hawkins2009, Li2022, Blanusa2022, Mankin2020}. Importantly, the effectiveness of an ensemble approach is closely tied to its size. Larger ensembles encompass a wider spectrum of structural differences across models and enable more precise statistical evaluation of projected changes in wind speed and power density.

The literature increasingly underscores the value of large ensemble datasets in energy-related climate impact studies. For example, \citet{Pryor2020}
reviewed multiple ensemble-based studies on wind energy, highlighting the role of model spread and regional variability in projected wind speed changes, including over North America.  \citet{Tobin2014} assessed wind energy changes in Europe using high-resolution ensemble simulations, demonstrating that ensemble size significantly influences the robustness of regional projections. In the Southern Hemisphere, \citet{Yu2021} analyzed future surface wind projections using a 50-member ensemble and found consistent hemispheric asymmetries in projected wind changes, confirming the critical role of internal variability and large ensemble sizes in identifying robust signals.

Recent advances in ensemble design have also improved the characterization of uncertainty. For instance, \citet{Lehner2020} demonstrated how combining large single-model ensembles with multi-model approaches provides a more nuanced decomposition of projection uncertainty, distinguishing between internal variability and model response differences. In the context of wind energy, this approach has proven invaluable in identifying regions where climate signals are strong and consistent across models, as well as areas where uncertainty remains high and further research is needed \citep{Gonzalez2020}. In addition, studies such as \citet{Moemken2018} and \citet{Karnauskas2018} have shown that changes in wind energy potential are not uniform across continents or seasons, reinforcing the need for geographically and temporally explicit assessments based on diverse model ensembles.

This paper builds on this growing body of literature by employing an extensive and high-resolution ensemble of 21 GCM--RCM combinations from the EURO-CORDEX initiative \citep{jacob2020eurocordex} to assess the future potential of wind energy under climate change. While previous studies have demonstrated the value of multi-model ensembles, our analysis goes further in several critical ways. First, we implement a formal robustness assessment framework following the IPCC AR6 `Approach C' to explicitly quantify where projected changes in wind speed and power are statistically significant and consistent across models. This formalized uncertainty treatment, rarely applied in wind energy assessments, allows for a much clearer interpretation of climate signals versus natural variability. Second, we extend the analysis beyond traditional mean-based approaches by introducing an event-based framework that quantifies the frequency and persistence of both high- and low-wind episodes, derived from ERA5 percentile thresholds. This temporal characterization is particularly valuable given the operational thresholds of wind turbines and the need for continuous generation. Third, we examine the sensitivity of projections to ensemble size, revealing that robust conclusions can shift or even invert when the number of models is insufficient. This finding highlights a critical risk in current practices where smaller or arbitrarily chosen subsets may lead to misleading energy planning outcomes. Moreover, the study delivers a spatially explicit, high-resolution view of wind power evolution across Europe, uncovering region-specific trends--such as robust future decreases in the North Atlantic and Mediterranean, and localized increases near the Iberian Peninsula. Taken together, these methodological advances make this study one of the most statistically robust and practically informative assessments of future wind energy potential under climate change. These findings are highly relevant for decision-makers aiming to integrate long-term climate resilience into energy infrastructure planning.

The paper is organized as follow. In Section~\ref{methods}, we outline the ensemble design, data sources, and key statistical metrics used to assess wind speed and power. Section~\ref{results} presents projected changes in wind conditions and evaluates their robustness using the IPCC AR6 framework. In Section~\ref{discussion}, we examine how ensemble size influences the reliability of results. Finally, Section~\ref{conclusions} summarizes the main insights and their relevance for wind energy planning under climate change.

\section{Methods and data}
\label{methods}

    \subsection{Multi-model ensemble of wind climate simulations}

    Given the enormous complexity of the climate system and the resulting uncertainties affecting simulations that aim to describe its evolution, the best approach to gain insights into future trends of atmospheric variables—specifically, in the case of this work, wind availability in Europe over the coming decades—is to consider a large multi-model ensemble of regional climate models that allow a comprehensive statistical analysis about its projected evolution \citep{Deser2012, Mankin2020}. \medskip

    To this purpose, the present work uses a large multi-model ensemble of wind speed over Europe. More specifically, the ensemble comprises twenty-one RCM-GCMs combinations from the EURO-CORDEX (Coordinated Regional Climate Downscaling Experiment) framework \citep{EUROCORDEX14, EUROCORDEX20} providing wind velocity components with 0.11$^\circ$ ($\approx$12.5 km) spatial resolution and 6-h temporal resolution for the period 1970 until 2005 for the historical simulations and for 2006 until 2100 for the RCP8.5 simulations. Table \ref{tab:cordex_models} presents the RCM-GCMs combinations comprising the ensemble and the notation used throughout this work. Additional details of the definition and performance of the different RCMs can be found on \citet{CLMmod} for the CLM-Community CCLM4-8-17 model, \citet{RCA4} for the Rossby Centre RCA4 model,  \citet{HIRHAM5} for the Danish Climate Centre HIRHAM5 model, \citet{crCLIM} for the COSMO-CLM accelerated version COSMO-crCLIM-v1-1, and, \citet{RACMO} for the Royal Netherlands Meteorological Institute (KNMI) RACMO22E model. \medskip

    \begin{table}
    \caption{EURO-CORDEX RCM and driving GCM combinations and notation used.}
    \label{tab:cordex_models}
    \resizebox{\textwidth}{!}{
        \begin{tabular}{r|ccccc}
        \toprule
        \multicolumn{1}{l}{} & \multicolumn{5}{c}{\textbf{RCM}} \\
        \cmidrule{2-6}
        \multicolumn{1}{c}{\textbf{GCM}} & CCLM4-8-17 & RCA4 & HIRHAM5 & COSMO-crCLIM-v1-1 & RACMO22E \\  \midrule
         CCCma-CanESM2 & \mone & & & &\\[-\jot]
         & \multicolumn{5}{c}{}\\[-\jot]
         MIROC-MIROC5 & \mtwo & & &  &\\[-\jot]
         & \multicolumn{5}{c}{}\\[-\jot]
         MPI-M-MPI-ESM-LR & & \mthree & \mtwel &  & \\[-\jot]
         & \multicolumn{5}{c}{}\\[-\jot]
         NCC-NorESM1-M & & \mfour & \mten & \msixt & \\[-\jot]
         & \multicolumn{5}{c}{}\\[-\jot]
         CNRM-CERFACS- & & \multirow{2}{*}{\mfive} & \multirow{2}{*}{\mthirt} & & \multirow{2}{*}{\meighteen} \\
         CNRM-CM5 & &  &  &  &\\[-\jot]
         & \multicolumn{5}{c}{}\\[-\jot]
         IPSL-IPSL-CM5A-MR & & \msix & \mfourt &  & \mtwenty \\[-\jot]
         & \multicolumn{5}{c}{}\\[-\jot]
         MOHC-HadGEM2-ES & & \mseven & \melev & \msevent & \mtwentyone \\[-\jot]
         & \multicolumn{5}{c}{}\\[-\jot]
         ICHEC-EC-EARTH & & \meight & \mnine & \mfivet  & \mnineteen \\
         \bottomrule
        \end{tabular}}          
    \end{table}
    
    \subsection{Wind energy statistics}
    
   Based on historical and future RCP8.5 simulations from the RCM-GCMs listed in Table \ref{tab:cordex_models}, various metrics were used in order to assess the potential evolution of wind resources over Europe. The analysis considers the 'mid-century' period from 2034 until 2060 and the 'end-of-century' period, covering from 2074 until 2100. Future changes are evaluated relative to the historical baseline period selected between 1979 and 2005. \medskip
    
    Firstly, we focused on a series of basic statistics regarding the wind speed at 10 meters height ($V_w$) and the corresponding wind power ($P_{wind}$), defined, per unit swept area, as: 
    \begin{equation}
        P_{wind} = \frac{\rho_{air}}{2} V_w^3,
        \label{eq:wind_energy}
    \end{equation}
    with $\rho_{air}$ being the air density.   In the present work, wind energy has been calculated based on the 10 meters wind simulations. The height of modern wind rotors can reach 120-150 meters, but with the advancement of technology, this value is expected to rapidly increase in near future. However, the corresponding wind power can easily be recalculated for the actual height of the rotor blades, once determined, by applying the widely used logarithmic relationship that describes the vertical distribution of the horizontal mean wind speed in the lowest portion of the planetary boundary layer, assuming a neutral stability profile \citep{Oke}.\medskip

    The selected basic statistics, for $V_w$ and $P_{wind}$, comprised the seasonal mean ($\overline{V_w}$,  $\overline{P_{wind}}$) and the annual mean of seasonal maxima ($\overline{\max(V_w)}$,  $\overline{\max(P_{wind})}$). \medskip
    
    Nevertheless, when analyzing wind potential for energy exploitation in a specific region, it is essential to consider not only the average wind speed and/or power values over various timescales (annual and seasonal) but also the stability of the resource itself \citep{FERRARI2020}. Additionally, considering that wind energy extraction is only possible within a specific range of wind intensity values, due to the inactivity of turbines under weak wind conditions, typically below 3 m/s, or during extreme winds, above 25 m/s \citep{sarkar2012wind}, this work also analyzes the future changes in upper and lower limits of wind events. In order to have consistent thresholds for both historical and future projections, we have used the ERA5 reanalysis data \citep{era5} as the baseline and established the 25th percentile (T25) and 75th percentile (T75) thresholds to define wind speed events and analyzed the frequency and average duration of events associated with these thresholds for a minimum set duration. More specifically, we analyzed on events with minimum duration of three, four, five and eight consecutive days, in which the daily maximum wind speed never reached T25 or exceeded T75.

    For all statistics and indices used in this work we have analyzed the ensemble mean of the historical conditions ($1979-2005$) and ensemble mean of the future changes between mid-century/end-of-century and historical. Furthermore, the ensemble robustness is analyzed following the Intergovernmental Panel on Climate Change (IPCC) Sixth Assessment Report \citep[AR6,][]{IPCC_2021_WGI_Atlas}. More specifically, we adopted the advanced alternative, \textit{approach C}, dividing robustness or uncertainty information into three categories: "robust change", "no change or no robust change", and "conflicting signal". This approach considers both the level of agreement among the models of the ensemble and whether the climate change signal emerges from internal variability or not. Ensemble agreement is considered when at least 80\% of the models agree on the sign of the change whereas a significant change is considered when at least 66\% of the models show a change greater than the variability threshold, which is defined as:
    \begin{equation}\label{eq:1}
        \gamma = \sqrt{2/20} \cdot 1.645 \cdot \sigma_{1yr},
    \end{equation}
    where $\sigma_{1yr}$ is the interannual standard deviation for the historical period. 
    
\section{Results}
\label{results}

    This section provides a detailed analysis of the future evolution of wind and the corresponding wind power potential under a changing climate. Wind speed and power conditions were analyzed, for each RCM-GCM presented on Table \ref{tab:cordex_models}, for historical period ($1979-2005$) and under RCP8.5 for \textit{Mid-century} period, ranging from 2034 until 2060 and for \textit{End-of-century} from 2074 to 2100. More specifically, the results for historical seasonal mean ($\overline{V_w}$,  $\overline{P_{wind}}$) and annual mean of seasonal maxima  ($\overline{\max(V_w)}$,  $\overline{\max(P_{wind})}$) are presented as well as the expected changes for future periods. Finally, to study the evolution of the duration of the operating periods of a hypothetical wind power plant, we analyzed the frequency and duration of events under the T25 and over the T75 thresholds. More specifically, we define the events:
    \begin{itemize}
        \item $X^{\mathrm{over}}_d$: events in which daily maximum wind speed exceeded T75 for at least $d$ consecutive days.
        \item $X^{\mathrm{under}}_d$: events in which daily maximum wind speed never reached T25 for at least $d$ consecutive days.
    \end{itemize}
    
    Following the definition of $X^{\mathrm{over}}_d$ and $X^{\mathrm{under}}_d$ events, two main indices were studied:
    \begin{itemize}
        \item $N_X$: annual mean number of $X^{\mathrm{over}}_d$ or $X^{\mathrm{under}}_d$ events, analyzed separately.
        \item $D_X$: annual average of days belonging to $X^{\mathrm{over}}_d$ or $X^{\mathrm{under}}_d$ events, analyzed separately.
    \end{itemize}
    
\begin{figure}
        \centering
        \includegraphics[width=0.49\linewidth]{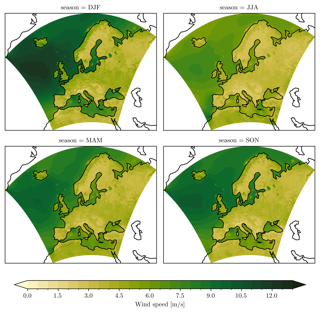}
        \includegraphics[width=0.49\linewidth]{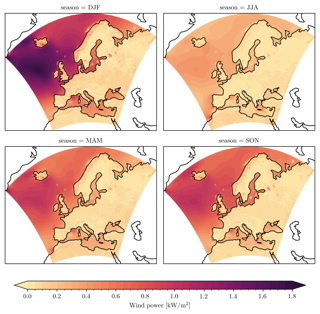}
        \includegraphics[width=0.49\linewidth]{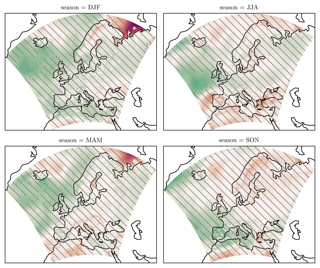}        
        \includegraphics[width=0.49\linewidth]{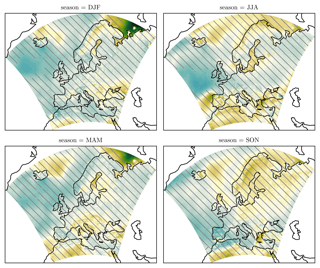}
        \includegraphics[width=0.49\linewidth]{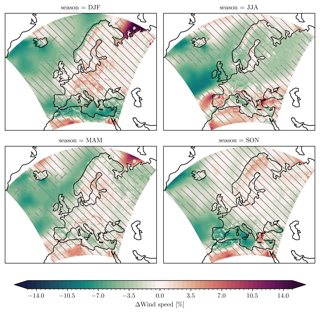}
        \includegraphics[width=0.49\linewidth]{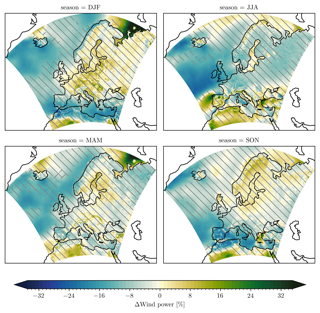}
        \caption{Ensemble-mean of $\overline{V_w}$ (left) and $\overline{P_{wind}}$ (right) for the historical period ($1979-2005$, top-row) and projected changes between future RCP8.5 and historical, for mid-century ($2034-2060$, middle-row) and end-of-century ($2074-2100$, bottom-row). Areas depicting a robust change are left without any overlay and areas depicting no robust change are represented with reverse-diagonal overlay.}
        \label{fig:means}
    \end{figure}

\begin{figure}
        \centering
        \includegraphics[width=0.49\linewidth]{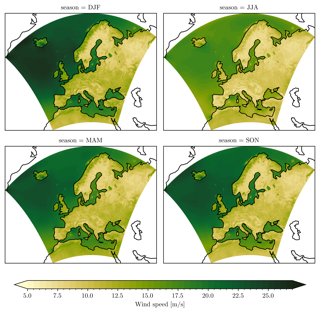}
        \includegraphics[width=0.49\linewidth]{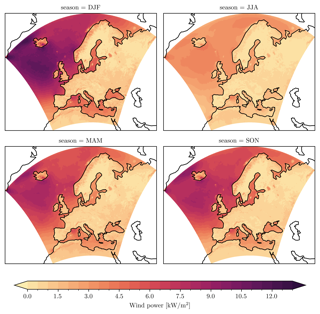}
        \includegraphics[width=0.49\linewidth]{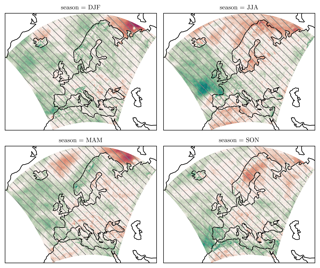}
        \includegraphics[width=0.49\linewidth]{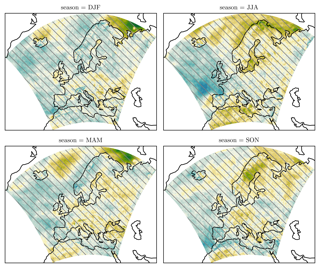}
        \includegraphics[width=0.49\linewidth]{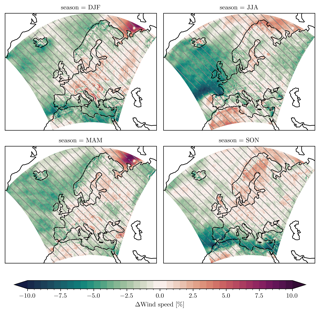}
        \includegraphics[width=0.49\linewidth]{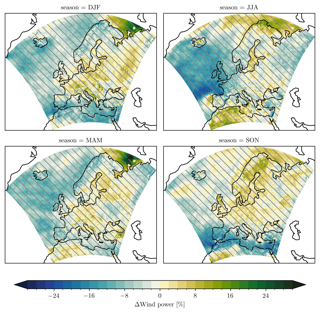}
        \caption{Ensemble-mean of  $\overline{\max(V_w)}$ (left) and $\overline{\max(P_{wind})}$ (right) for the historical period ($1979-2005$, top-row) and projected changes between future RCP8.5 and historical, for mid-century ($2034-2060$, middle-row) and end-of-century ($2074-2100$, bottom-row). Areas depicting a robust change are left without any overlay and areas depicting no robust change are represented with reverse-diagonal overlay.}
        \label{fig:max}
    \end{figure}

    As introduced in \S\ref{methods}, the robustness and uncertainty of the results across the different GCM-RCMs is assessed using the IPCC AR6 approach C, defining three different categories as follows:
    \begin{enumerate}
    \item Robust change (significant change and high model agreement):
        \begin{itemize}
            \item[] $\geq 80\%$ of models agree on the sign of change, and,
            \item[] $\geq 66\%$ of models present a change greater than the variability threshold ($\gamma$).                
        \end{itemize} 
    \item No robust change: 
         \begin{itemize}
            \item[] $<66\%$ of models present a change greater than the variability threshold.             
        \end{itemize} 
    \item Conflicting signal (significant change but low agreement): 
        \begin{itemize}
            \item[] $<80\%$ of models agree on the sign of change, and, 
            \item[] $\geq 66\%$ of models present a change greater than the variability threshold.       
        \end{itemize} 
    \end{enumerate}     
    Therefore, for all results presented henceforward relating future projected changes, additional hatched/overlay areas have been included. Following the defined above-mentioned categories, areas depicting a robust change are left without any overlay, areas depicting no robust change are represented with \textit{reverse-diagonal} overlay, and, areas depicting conflicting signal are represented with \textit{crossed-lines} overlay. Figures  \ref{fig:means} and \ref{fig:max} present the results for the seasonal mean and annual mean of seasonal maxima, respectively, for wind speed (left) and wind power (right). Each figure presents, on the top row, the historical ensemble-mean for the period $1979-2005$, on the middle-row, the ensemble-mean of projected changes between mid-century ($2034-2060$) and the historical period, and, on the bottom-row, the ensemble-mean of projected changes for end-of-century conditions ($2074-2100$). \medskip

    Regarding the seasonal mean (Fig. \ref{fig:means}), the largest values for both wind speed and wind power are observed in the North-Atlantic Ocean and the Norwegian Sea with values up to 13 m/s and 1.8 kW/m$^2$ in Winter (DJF). Future projections for this region during Winter, present robust decreases of $\approx 5-8\%$ and $\approx 10-15\%$, for wind speed and wind power, respectively. Areas with robust decreases can be identified in the Atlantic Ocean for all the remaining seasons, with the largest decreases depicted for Summer (JJA) for the West European basin area charcterized by a robust decreases of up to -12\%. It can also be highlighted for Summer, a robust increase of $\approx 10\%$ for end-of-century wind speed and $\approx 25\%$ for wind energy, in the nearshore region of Spain and Portugal. This variation in the projected change signal for summer in the Atlantic Ocean, could be due to possible changes in predominant wind directions and synoptic circulations \citep{herrera2023projected}. Regarding the Mediterranean Sea, where historical seasonal mean wind speeds of 3-8 m/s are depicted, future projections at the end-of-the-century show a general robust decrease in the central and western basin for Winter and Fall. Additional robust changes for end-of-century conditions are depicted for Spain with a robust decrease of $\approx -10\%$ during Fall (SON) and a robust increase of $\approx 5\%$ during Summer (JJA). \medskip

    Regarding the annual means of seasonal maxima results (Fig. \ref{fig:max}), as expected and consistently with previous works\citep{laurila2021climatology}, the largest historical values for both wind speed and wind power are observed in the North-Atlantic Ocean and the Norwegian Sea with values up to 25 m/s and 12 kW/m$^2$ in Winter. Future projections for this region, present small non-robust decreases of $\approx 10\%$ with the largest decreases depicted for Summer (JJA) in the Atlantic Ocean-Bay of Biscay region for mid-century conditions and robust decreases of under 10\% for end-of-century. The largest future changes are observed for the Barents Sea with increases of $\gtrapprox 10\%$ and  $\gtrapprox 20\%$ for wind speed and wind power, respectively, for Winter (DLF) and Spring (MAM) for both future periods although with no robustness. For both seasonal mean and maxima, future changes over land present a larger spatial variability than over sea. This could be due to the different way in which various models parametrize the complex interaction between atmosphere and land.  Moreover, the different land use evolution projected by the various models, in situations not characterized by a clear signal in wind trend, can contribute to wind projections that are highly variable from one area to another.\medskip

\begin{figure}
        \centering
        \includegraphics[width=0.49\linewidth]{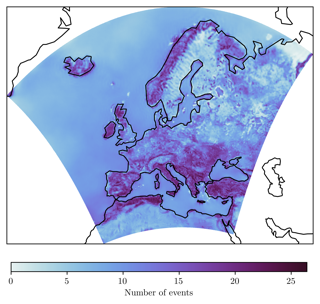}
        \includegraphics[width=0.49\linewidth]{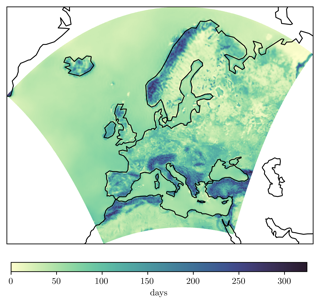}
        \includegraphics[width=0.49\linewidth]{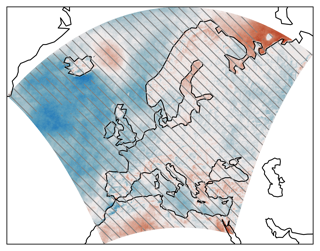}
        \includegraphics[width=0.49\linewidth]{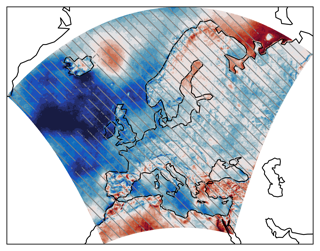}
        \includegraphics[width=0.49\linewidth]{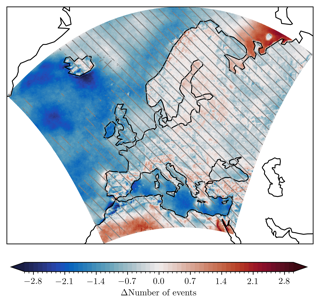}
        \includegraphics[width=0.49\linewidth]{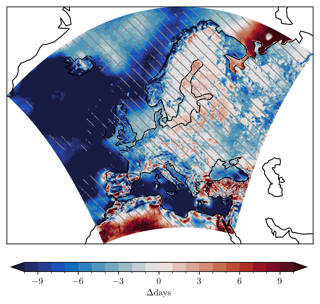}
        \caption{Ensemble-mean of $N_X$ (left) and $D_X$ (right) for $X^{\mathrm{over}}_4$ events. Historical period ($1979-2005$, top-row) and projected changes between future RCP8.5 and historical, for mid-century ($2034-2060$, middle-row) and end-of-century ($2074-2100$, bottom-row). Areas depicting a robust change are left without any overlay and areas depicting no robust change are represented with reverse-diagonal overlay.}
        \label{fig:Over75-4}
    \end{figure}

\begin{figure}
        \centering
        \includegraphics[width=0.49\linewidth]{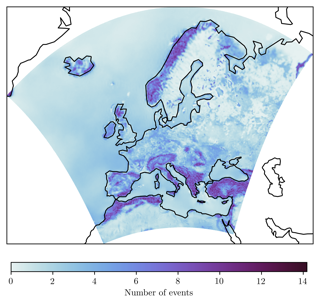}
        \includegraphics[width=0.49\linewidth]{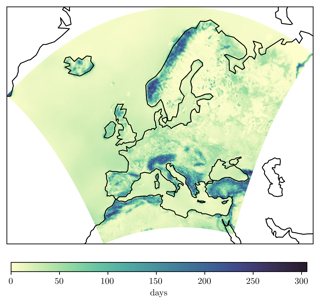}
        \includegraphics[width=0.49\linewidth]{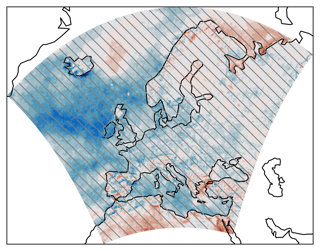}
        \includegraphics[width=0.49\linewidth]{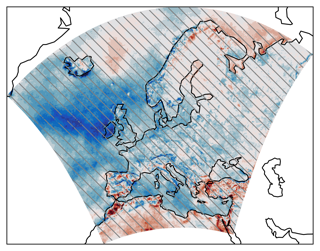}
        \includegraphics[width=0.49\linewidth]{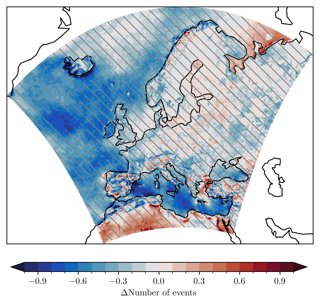}
        \includegraphics[width=0.49\linewidth]{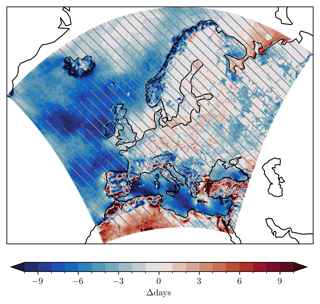}
        \caption{Ensemble-mean of $N_X$ (left) and $D_X$ (right) for $X^{\mathrm{over}}_8$ events. Historical period ($1979-2005$, top-row) and projected changes between future RCP8.5 and historical, for mid-century ($2034-2060$, middle-row) and end-of-century ($2074-2100$, bottom-row). Areas depicting a robust change are left without any overlay and areas depicting no robust change are represented with reverse-diagonal overlay.}
        \label{fig:Over75-8}
    \end{figure}

    Moving on to analyze the possible evolution of the tails of the wind intensity distribution, Figures \ref{fig:Over75-4} and \ref{fig:Over75-8} show the results regarding the $X^{\mathrm{over}}_4$ and $X^{\mathrm{over}}_8$, i.e., the events where the daily maxima exceeds T75 for a minimum of 4 and 8 days, respectively. For brevity, the results related to events lasting at least three and five days are not shown, but they reported similar results to those for events lasting at least four days. Each figure presents, $N_X$ and  $D_X$ on the left and right column panels, respectively, and, from top to bottom rows, the historical ensemble-mean ($1979-2005$), projected changes for mid-century ($2034-2060$) and end-of-century conditions ($2074-2100$). Starting from the analysis of the shorter events ($X^{\mathrm{over}}_4$) depicted in Figure \ref{fig:Over75-4}, the most relevant result is the significant and robust decreasing both in the number of the events and in the annual average duration (i.e., less events and shorter) expected for the mid of the century in the North-West Atlantic Ocean. At the same time a robust increase both in the number of events and their duration is observed in a portion of the Barents Sea. This trend becomes even more pronounced toward the end of the century, when the substantial decrease in both the number of events and their duration is expected to extend to a large part of the Atlantic Ocean. Moreover, by the end of the century, the decrease in the number of events and their duration also becomes robust across the entire Mediterranean basin. This robust decrease on the number of events and their duration is coherent with the projected mean sea level pressure increase, during Winter, extending from North Atlantic to the European region described in \citep{giorgi2007european} and also highlighted by a projected increase NAO's positive phase and a northward shift of the Atlantic storm track \citep{coppola2005bimodality, giorgi2008climate}. 
    Once again, conclusions regarding land areas remain neither robust nor univocal. 
      
    The same type of trend, though less pronounced in magnitude, can be observed in the analysis of the projected evolution of longer events (Figure \ref{fig:Over75-8}). In this case, however, the projected signal for mid-century is not robust, whereas it becomes robust toward the end of the century, particularly for the Mediterranean Basin and small portions of the Atlantic Ocean. \medskip

\begin{figure}
        \centering
        \includegraphics[width=0.49\linewidth]{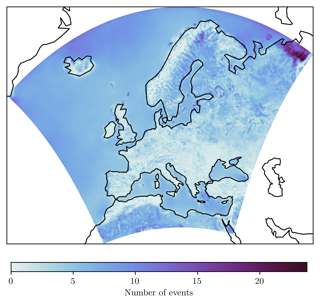}
        \includegraphics[width=0.49\linewidth]{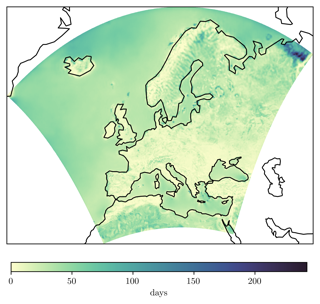}
        \includegraphics[width=0.49\linewidth]{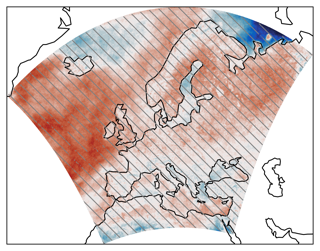}
        \includegraphics[width=0.49\linewidth]{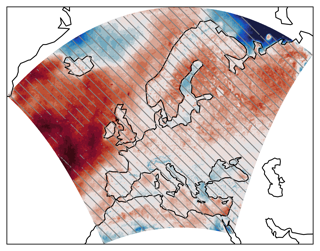}
        \includegraphics[width=0.49\linewidth]{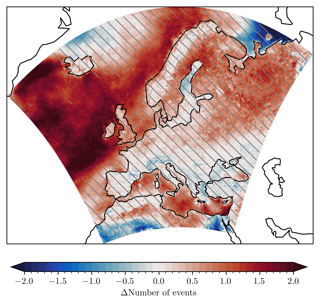}
        \includegraphics[width=0.49\linewidth]{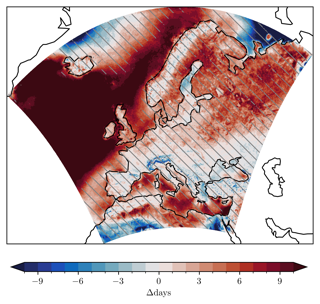}
        \caption{Ensemble-mean of $N_X$ (left) and $D_X$ (right) for $X^{\mathrm{under}}_4$ events. Historical period ($1979-2005$, top-row) and projected changes between future RCP8.5 and historical, for mid-century ($2034-2060$, middle-row) and end-of-century ($2074-2100$, bottom-row). Areas depicting a robust change are left without any overlay and areas depicting no robust change are represented with reverse-diagonal overlay.}
        \label{fig:Under25-4}
    \end{figure}

\begin{figure}
        \centering
        \includegraphics[width=0.49\linewidth]{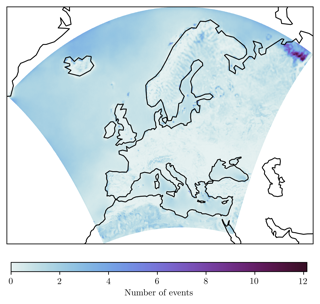}
        \includegraphics[width=0.49\linewidth]{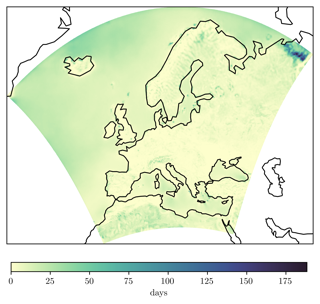}
        \includegraphics[width=0.49\linewidth]{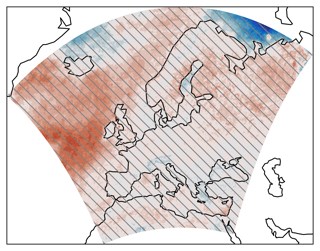}
        \includegraphics[width=0.49\linewidth]{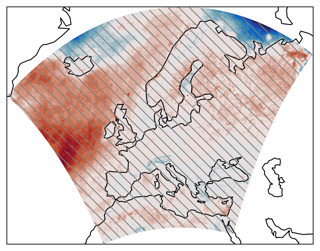}
        \includegraphics[width=0.49\linewidth]{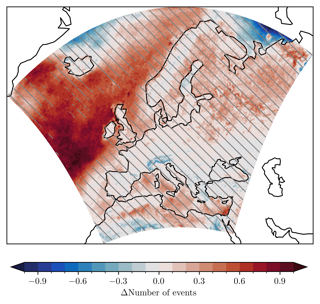}
        \includegraphics[width=0.49\linewidth]{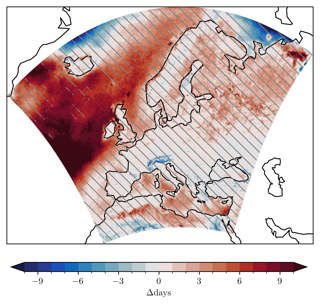}
        \caption{Ensemble-mean of $N_X$ (left) and $D_X$ (right) for $X^{\mathrm{under}}_8$ events. Historical period ($1979-2005$, top-row) and projected changes between future RCP8.5 and historical, for mid-century ($2034-2060$, middle-row) and end-of-century ($2074-2100$, bottom-row). Areas depicting a robust change are left without any overlay and areas depicting no robust change are represented with reverse-diagonal overlay.}
        \label{fig:Under25-8}
    \end{figure}

    Moving on to analyze events characterized by calm wind conditions, Figures \ref{fig:Under25-4} and \ref{fig:Under25-8} show the results for the $X^{\mathrm{under}}_4$ and $X^{\mathrm{under}}_8$, i.e., the events where the daily maxima does not reach T25 for a minimum of 4 and 8 days, respectively. Regarding the projected changes in $N_X$ (left) and $D_X$ (right) for mid-century conditions, no robust variations are projected for either shorter ($X^{\mathrm{under}}_4$, Figure \ref{fig:Under25-4}, middle-row) or longer events ($X^{\mathrm{under}}_8$, \ref{fig:Under25-8}). Moving on to the analysis of the projected evolution toward the end of the century, a significant increase in $N_X$ and $D_X$ is expected in the Atlantic Ocean and the Mediterranean Sea for the shorter events ($X^{\mathrm{under}}_4$), but only in the Atlantic Ocean for the longer events ($X^{\mathrm{under}}_8$).\\
    In summary, the results from the large ensemble analyzed in this study show a general weakening trend of the wind, given by a projected increase of lower-intensity events and a projected decrease in high-intensity events, especially towards the end of the century, in the Atlantic Ocean and the Mediterranean Basin, while no robust conclusions can be drawn regarding land areas.

\section{Discussion - the role of the ensemble}
\label{discussion}

Despite the results presented in the previous section not indicating a clear and distinct trend in the evolution of wind patterns, either for the near future or the end of the century, with the exception of a few limited marine areas, some significant conclusions can still be drawn. A key strength of this work is the large number of members (RCM-GCM) used to create the analyzed ensemble. We will demonstrate how using an ensemble with a smaller number of members can lead to contradictory conclusions, even when those conclusions appear to be robust. For this we have selected different 5-member ensembles and reproduced the different indices presented in the previous sections of this work. Table \ref{tab:sub_ensembles} presents the different sub-ensembles that have been selected to analyze the seasonal maxima and the over-threshold wind events. \medskip

\begin{table}
    \caption{RCM-GCMs combinations comprising the sub-ensembles exploited to analyse different indeces ({\textbf{Winter $\overline{\max(P_{wind})}$}}, {\textbf{Spring $\overline{\max(P_{wind})}$}} and {\textbf{$X^{\mathrm{over}}_4$}}}
    \label{tab:sub_ensembles}
        \begin{tabular}{cc}
        \toprule
        \multicolumn{2}{c}{\textbf{Winter $\overline{\max(P_{wind})}$}} \\
        \midrule
        A & B \\
        \mseven     & \mfourt      \\   
        \meight     & \mtwenty     \\   
        \mnine      & \mtwentyone  \\   
        \mfivet     & \mseven      \\   
        \mnineteen  & \msix        \\   
        \midrule
        \multicolumn{2}{c}{\textbf{Spring $\overline{\max(P_{wind})}$}} \\
        \midrule
        A & B \\
        \mtwentyone & \mtwenty  \\   
        \msevent    & \meight   \\   
        \mseven     & \mfive    \\   
        \mone       & \mfour    \\   
        \mtwo       & \msix     \\ 
        \midrule
        \multicolumn{2}{c}{\textbf{$X^{\mathrm{over}}_4$}} \\
        \midrule
        A & B \\
        \mthree     & \mseven   \\   
        \mtwenty    & \mnine    \\   
        \mnineteen  & \melev    \\   
        \meighteen  & \mfivet   \\   
        \msix       & \msevent  \\           
        \bottomrule
        \end{tabular}
\end{table}

\begin{figure}
    \centering
    \includegraphics[width=\linewidth]{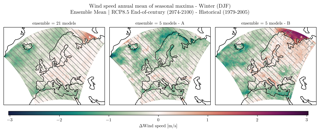}
    \includegraphics[width=\linewidth]{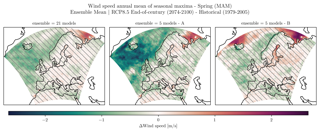}
    \includegraphics[width=\linewidth]{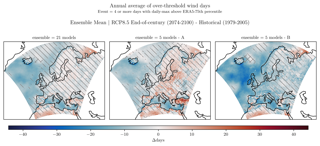}
    \caption{Ensemble-mean of projected changes between end-of-century RCP8.5 and historical for Winter $\overline{\max(P_{wind})}$ (top-row), Spring $\overline{\max(P_{wind})}$ (middle-row) and $D_X$ for $X^{\mathrm{over}}_4$ events (bottom-row). Each row presents from the left to right, the 21-members ensemble and two different ensemble of 5-members each enumerated in Table \ref{tab:sub_ensembles}. Areas depicting a robust change are left without any overlay, areas depicting no robust change are represented with reverse-diagonal overlay, and, areas depicting conflicting signal are represented with crossed-lines overlay.}
    \label{fig:ensembles-comparison}
\end{figure}

Figure \ref{fig:ensembles-comparison} shows the evolution of some of the indices examined in the previous sections, obtained from the ensembles with different numbers of members. Specifically, it depicts, in the first two rows, the ensemble-mean changes for end-of-century for $\overline{\max(P_{wind})}$ during Winter (DJF) and Spring (MAM), and, on the bottom-row, the  ensemble-mean changes for end-of-century of $D_X$ for $X^{\mathrm{over}}_4$. In each row, the first column refers to the projected variations based on the entire ensemble composed of 21 members whereas the second and third column refer to the to the projected variations based on the A and B sub-ensembles, respectively, each one composed of the 5 different RCM-GCM enumerated in Table \ref{tab:sub_ensembles}. \medskip

From the Winter and Spring $\overline{\max(P_{wind})}$, it can be observed that the projected variations are not robust across the entire domain for the ensemble considering 21 RCM-GCMs. 
These conclusions, however, are reversed if the analysis is based on ensembles composed of a reduced number of members. We can even identify areas where the variations are robust but have opposite signs depending on the ensemble considered. This is the case, for example, for the Spring  $\overline{\max(P_{wind})}$ (middle-row) where the maritime area north of Iceland, presents, a robust signal of either an increase or a decrease of the wind speed annual mean of  seasonal maxima, depending on the sub-ensemble. A similar behavior can be observed for the significant and, at times, robust increase in the Winter $\overline{\max(P_{wind})}$ (first row) between the Barents Sea, Finland, and northwestern Russia. This increase is contradicted by a robust decrease when referring to a different ensemble. \medskip

A more or less similar conclusion can be drawn by analyzing $D_X$ for $X^{\mathrm{over}}_4$ events (third row). In this case, while it is true that a large area encompassing the North Atlantic and the Mediterranean shows a robust decrease in $D_X$ both in the analysis of the full ensemble and in the analysis of the two sub-ensembles, other areas present diametrically opposite conclusions. For example, the area of the Anatolian Peninsula presents a robust increase in $D_X$ or a decrease, depending on the sub-ensemble considered. Similar conclusions can be drawn by analyzing the area around the Baltic Sea. In contrast, the analysis of the full ensemble clearly indicates that the projected evolution of $D_X$ for $X^{\mathrm{over}}_4$ events by the end of the century is characterized by low magnitude and lack of robustness for both areas. From this example, it becomes evident that, when analyzing the possible long-term evolution of climate variables, it is essential to rely on an ensemble with the highest as possible number of members. This ensures that all potential sources of uncertainty inherent in such simulations are adequately accounted for.

\section{Conclusions}
\label{conclusions}
The increasing urgency of transitioning to renewable energy systems under a changing climate has placed wind energy at the forefront of sustainable energy strategies. However, the inherent variability of wind and the uncertainties introduced by climate change pose serious challenges for long-term planning. This study has addressed these challenges through a high-resolution, multi-model ensemble analysis of future wind energy potential across Europe. By leveraging 21 RCM--GCM combinations from the EURO-CORDEX initiative, we have delivered a detailed spatial and temporal assessment of projected wind speed and power under the RCP8.5 scenario. Our findings show that while large-scale, consistent changes in wind conditions are limited, specific regions--particularly parts of the Atlantic and Mediterranean--are projected to experience robust increases or decreases in wind energy potential. More importantly, we demonstrate that ensemble diversity is not just a methodological preference but a critical factor in avoiding misleading or contradictory conclusions: results derived from small sub-ensembles often diverge sharply, even reversing sign, compared to those from the full ensemble. Furthermore, our use of event-based metrics reveals a potential weakening of high-wind episodes and an increase in prolonged low-wind periods, with implications for grid reliability and energy storage needs. These insights underscore the importance of using large, carefully constructed ensembles and statistically rigorous frameworks--such as the IPCC AR6 robustness criteria--to inform energy policy.

Looking ahead, future work could apply this framework to alternative emission scenarios such as SSP-based pathways, and importantly, integrate bias correction techniques to improve the usability of climate projections for operational wind energy assessments. Extending the analysis toward impact-based studies - such as estimating actual energy production, grid reliability, or economic performance - would further bridge the gap between climate modeling and decision-making in the energy sector. As nations move toward decarbonization, robust climate-energy modeling like that presented here will be essential in guiding infrastructure investment and ensuring a reliable and climate-adapted energy future.

\section{Acknowledgements}
Project funded under the National Recovery and Resilience Plan (NRRP), Mission 4 Component 2 Investment 1.3 – Call for tender No. 1561 of 11.10.2022 of Ministero dell’Università e della Ricerca (MUR); funded by the European Union – NextGenerationEU. Award Number: Project Code code PE0000021, Concession Decree No. 1561 of 11.10.2022 adopted by Ministero dell’Università e della Ricerca (MUR), CUP D33C22001330002, Project title ``Network 4 Energy Sustainable Transition -- NEST''.

\bibliographystyle{elsarticle-harv} 
\bibliography{cas-refs}

\end{document}